 \documentclass[final,5p,times,twocolumn]{elsarticle}

\usepackage{amssymb,graphicx,dcolumn,amsmath,verbatim,rotating,bm}
\biboptions{comma,square,sort&compress}

\begin{document}

\begin{frontmatter}



\title{Nuclear charge radii of potassium isotopes beyond $N=28$}


\author[MPIK]{K.~Kreim\corref{cor1}}
\ead{kim.kreim@cern.ch}
\author[IKS]{M.~L.~Bissell}
\author[IKS]{J.~Papuga}
\author[MPIK]{K.~Blaum}
\author[IKS]{M.~De~Rydt}
\author[IKS]{R.~F.~Garcia~Ruiz }
\author[IAA]{S.~Goriely}
\author[IKS]{H.~Heylen}
\author[CERN]{M.~Kowalska}
\author[MPIK,KCHM]{R.~Neugart}
\author[IKS]{G.~Neyens}
\author[KCHM,IKD]{W.~N\"ortersh\"auser}
\author[IKS]{M.~M.~Rajabali}
\author[GSI,HIM]{R.~S\'{a}nchez~Alarc\'{o}n}
\author[DPNY]{H.~H.~Stroke}
\author[MPIK,CERN]{D.~T.~Yordanov}

\address[MPIK]{Max-Planck-Institut f\"ur Kernphysik, Saupfercheckweg 1, D-69117 Heidelberg, Germany}

\address[IKS]{Instituut voor Kern- en Stralingsfysica, KU Leuven, B-3001 Leuven, Belgium}

\address[IAA]{Institut d'Astronomie et d'Astrophysique, CP-226, Universit\'{e} Libre de Bruxelles, B-1050 Brussels, Belgium}

\address[CERN]{Physics Department, CERN, CH-1211 Geneva 23, Switzerland}

\address[KCHM]{Institut f\"ur Kernchemie, Universit\"at Mainz, D-55128 Mainz, Germany}

\address[IKD]{Institut f\"ur Kernphysik, Technische Universit\"at Darmstadt, D-64289 Darmstadt, Germany}

\address[GSI]{GSI Helmholtzzentrum f\"ur Schwerionenforschung GmbH, D-64291 Darmstadt, Germany}

\address[HIM]{Helmholtz-Institut Mainz, 55099 Mainz, Germany}

\address[DPNY]{Department of Physics, New York University, New York, NY 10003, USA}

\cortext[cor1]{Corresponding author}

\begin{abstract}
We report on the measurement of optical isotope shifts for $^{38,39,42,44,46\text{-}51}$K relative to $^{47}$K from which changes in the nuclear mean square charge radii across the $N=28$ shell closure are deduced. The investigation was carried out by bunched-beam collinear laser spectroscopy at the CERN-ISOLDE radioactive ion-beam facility. Mean square charge radii are now known from $^{37}$K to $^{51}$K, covering all $\nu f_{7/2}$-shell as well as all $\nu p_{3/2}$-shell nuclei. These measurements, in conjunction with those of Ca, Cr, Mn and Fe, provide a first insight into the $Z$ dependence of the evolution of nuclear size above the shell closure at $N=28$. 

\end{abstract}


\end{frontmatter}

% \linenumbers

%%\section{}
%%\label{}

Mean square charge radii of nuclei in the calcium region ($Z=20$) have been the subject of extensive investigation, both experimentally \cite{Touchard1982,Martensson-Pendrill1990,Behr1997,Charlwood10,Blaum2008,Avgoulea11} and theoretically \cite{Zamick1971,Talmi1984,Caurier01,Zamick10,Blaum2008,Avgoulea11}. Although existing data cover the full $\nu f_{7/2}$ orbital, very little is known on the nuclei above N=28 with valence neutrons in the $p_{3/2}$ orbital. 
Furthermore, substantial structural changes are predicted in this region, including the inversion and subsequent re-inversion of the $\pi s_{1/2}$ and $\pi d_{3/2}$ shell-model orbitals \cite{Touchard1982,Papuga12} and the development of subshell closures at $N=32$ and $N=34$ \cite{Honma2005,Gade2008,Sorlin08,Wienholtz2013a,Steppenbeck13}. Whilst theoretical models of charge radii across the $\nu f_{7/2}$ shell have been successful in describing the trend observed for calcium \cite{Talmi1984,Caurier01}, little is known about how this observable would be influenced by the anticipated structural evolution in the region beyond $N=28$.

The inversion of the odd-$A$ potassium ground-state configuration from $I=3/2$ ($\pi d_{3/2}$) in $^{39\text{-}45}$K to $I=1/2$ ($\pi s_{1/2}$) for $^{47}$K ($N=28$) was reported by \citet{Touchard1982} in 1982 and has subsequently been reproduced by nuclear shell-model and mean-field calculations \cite{Gade06,Otsuka05,Otsuka06}. The question of how the $\pi d_{3/2}$ orbital evolves whilst filling the $\nu p_{3/2}$ orbital has been addressed in a recent paper by \citet{Papuga12} for the even-$N$ isotopes $^{49}$K and $^{51}$K, having respectively spin $1/2$ and $3/2$. In the present work, the ground-state structure of the odd-$N$ isotopes $^{48}$K and $^{50}$K and a detailed analysis of the spin determination for $^{51}$K are presented. With the spin measurements on $^{48\text{-}51}$K the contradictory assignments found in decay spectroscopy data \cite{Krolas11,Broda10,Baumann98,Crawford09,Perrot06} are resolved.
These spins in combination with the magnetic moment of $^{48}$K fully define the region of inversion \cite{Papuga13}.

In this letter we report on the measurement of optical hyperfine structure and isotope shifts for $^{38,39,42,44,46\text{-}51}$K relative to $^{47}$K, from which changes in the nuclear mean square charge radii are deduced. The measurements were carried out at the collinear laser spectroscopy setup COLLAPS \cite{Neugart1981} at ISOLDE-CERN \cite{Kugler00}. Neutron-rich potassium isotopes were produced by $1.4$-GeV protons impinging onto a thick UC$_{x}$ target. The isotopes were surface ionized, accelerated to $40$\,keV, mass separated and directed to the ISOLDE cooler-buncher ISCOOL \cite{Franberg2008}. After ISCOOL the ion bunches were directed to the collinear laser spectroscopy beam line, where they were neutralized in collisions with potassium vapor. The atomic $4s\ ^{2}S_{1/2}\rightarrow 4p\ ^{2}P_{1/2}$ transition ($\lambda=769.9$\,nm) was excited by light generated from a cw titanium-sapphire ring laser. Frequency scanning was performed by applying an additional acceleration potential to the neutralization cell, thus varying the Doppler-shifted laser frequency in the reference frame of the atoms.
 A light collection system consisting of four separate photomultiplier tubes (PMTs) with imaging optics placed perpendicular to the beam path was used to detect the resonance fluorescence photons. Counts from the PMTs were only accepted during the time in which the atom bunches passed through the light collection system. The background from scattered photons was thus reduced by a factor of $\sim10^{4}$ \cite{Nieminen2003,Flanagan2009}. 

\begin{figure}[htb]
\begin{center}
\includegraphics[width=1.0\linewidth]{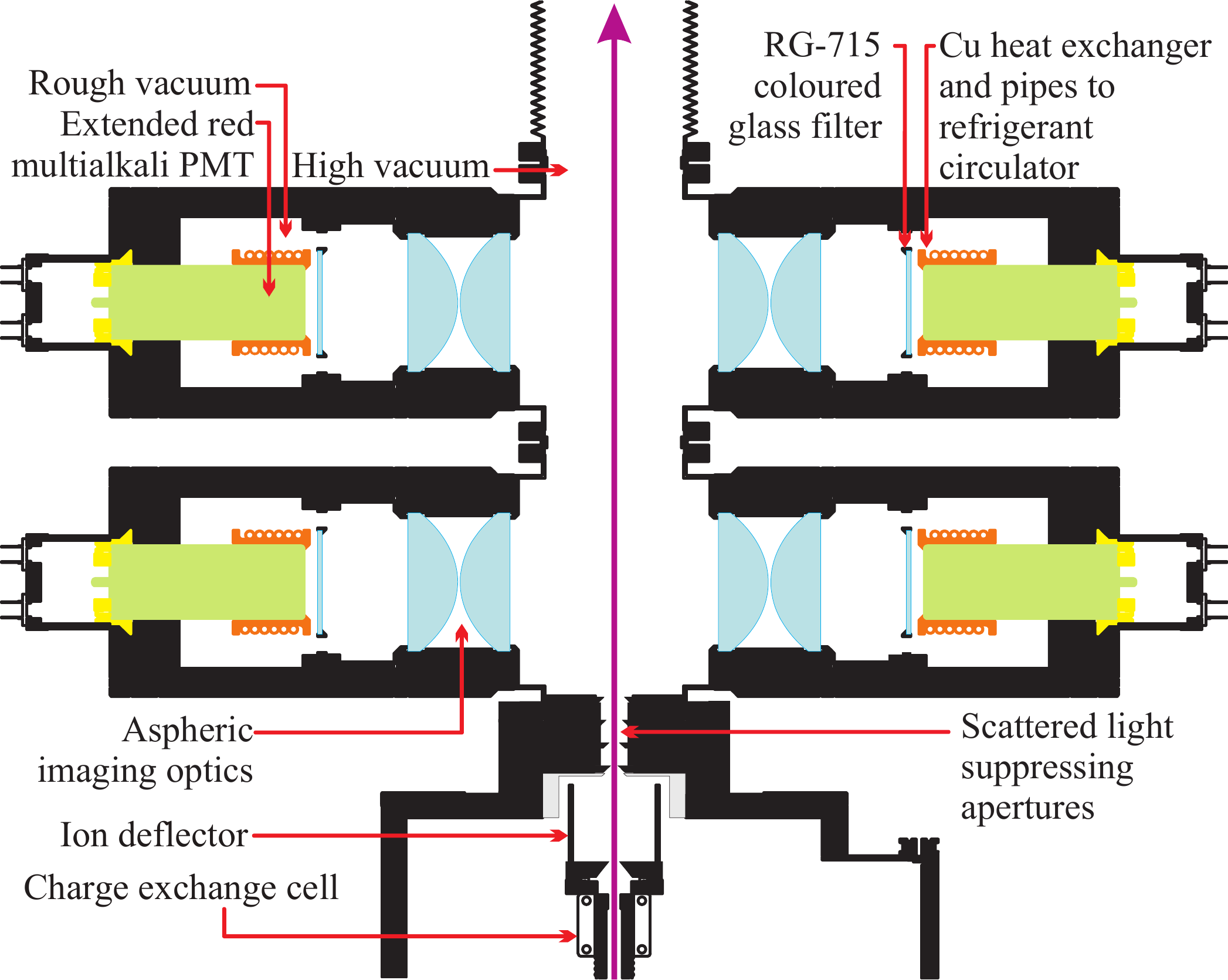}
\caption{\label{Detect} (Color online) Cut through the optical detection region (top view). The atom and laser beams enter the detection region through the charge exchange cell (bottom). For details see text.}
\end{center}
\end{figure}

Optical spectroscopy of K isotopes is hindered by the relatively slow decay rate of the atomic transition ($3.8 \times 10^7$\,s$^{-1}$) as well as a low (2.5\,\%) quantum efficiency of the PMTs with a high heat-related dark count rate.
In order to perform the measurements on a $^{51}$K beam of approximately $4000$\,ions/s, a new optical detection region was developed. Eight 100-mm diameter aspheric lenses were used to precisely image the fluorescence of the laser-excited K atoms onto four extended-red multialkali PMTs in the arrangement shown in Fig.~\ref{Detect}. The light-collection efficiency of this system is approximately twice that of the previous standard system described in \citet{Mueller1983}, whilst the background from scattered laser light is an order of magnitude lower. 
The PMTs were maintained at $-40\,^\circ$C using a refrigerant circulator to reduce dark counts and held under vacuum to prevent ice formation. RG715 colored glass filters were placed in front of the PMTs in order to cut the strong visible beam light originating from stable contaminant beams of Ti and Cr excited in collisions in the charge exchange cell.

\begin{figure}[h]
\begin{center}
  \includegraphics[width=0.98\linewidth]{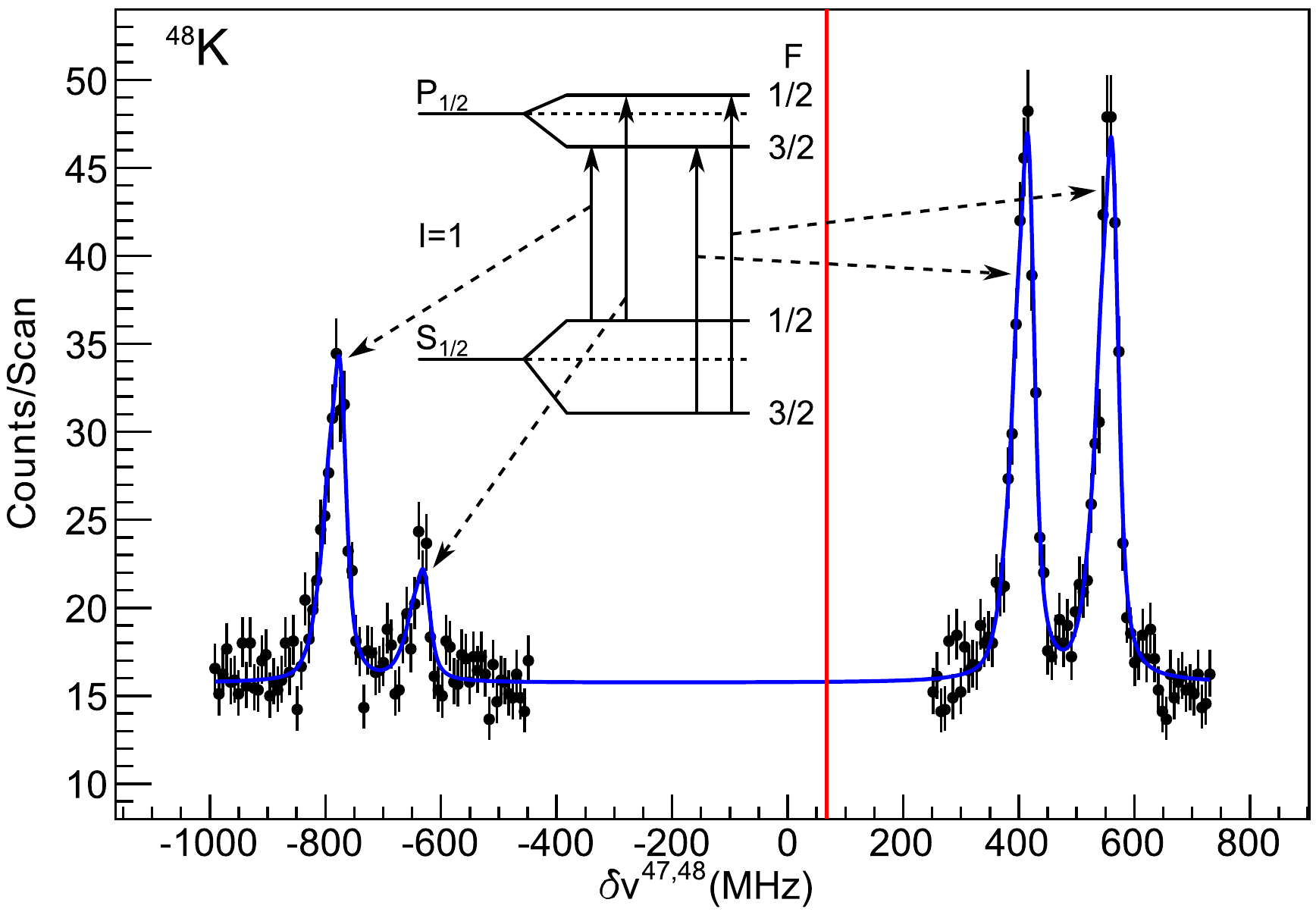}%
 \caption{(Color online) Optical spectrum of $^{48}$K. Also shown is the fitted hfs (solid line) and the hfs centroid (vertical line). The  ($S_{1/2}$) and ($P_{1/2}$) level scheme is shown as an inset. \label{48K}}
\end{center} 
\end{figure}
Isotope shift measurements were performed relative to $^{47}$K. 
The spectra were fitted with a hyperfine structure (hfs) using a line shape composed of multiple Voigt profiles \cite{Klose12, Bendali1981}, which were found to describe most adequately the asymmetric line shape, using $\chi^{2}$ minimization (see Fig.~\ref{48K} and \ref{50K}).  
For the fit a two-parameter first-order hyperfine splitting \cite{Kopfermann58} was used. The magnetic hyperfine parameters $A$ and the centroid of each hfs were extracted.
A sample spectrum for $^{48}$K is shown in Fig.~\ref{48K} together with the fitted hfs spectrum. Also shown are the energy levels of the ground state ($S_{1/2}$) and excited state ($P_{1/2}$) with allowed transitions.

In Table \ref{Results} we give the ground-state spins deduced from the measured hfs patterns. The spins measured for $^{37\text{-}47}$K, previously reported in \cite{Touchard1982,Koepf1969}, are confirmed by the present analysis. Our new spins measured for $^{49}$K and $^{51}$K have been published in \cite{Papuga12}. 
The procedure to determine the spins of $^{48}$K, $^{50}$K and $^{51}$K will be described in the following.

\begin{table}[th]%[H] %add [H] placement to break table across pages
 \caption{Spins, isotope shifts referenced to $^{47}$K ($\delta \nu^{47,A}$), and changes in mean square charge radii ($\delta \langle r^{2}\rangle^{47,A}$) of potassium isotopes in the atomic transition $4s\ ^{2}S_{1/2}\rightarrow 4p\ ^{2}P_{1/2}$. Statistical errors are shown in round brackets, systematic errors in square brackets. Complementary data $\delta \nu^{39,A}$ from \cite{Behr1997} and \cite{Touchard1982} (published without systematic errors), referenced to $^{39}$K, are quoted in the fourth column.}
\label{Results}
\centering\small
 \begin{tabular}{ccD{.}{.}{3.9}D{.}{.}{3.4}D{.}{.}{3.11}}
 $A$ & $I^{\pi}$ & \multicolumn{1}{c}{$\delta \nu^{47,A}$ (MHz)} & \multicolumn{1}{c}{$\delta \nu^{39,A}$ (MHz)}  & \multicolumn{1}{c}{$\delta \langle r^{2}\rangle^{47,A}$ (fm$^{2})$}\\
 \hline
 37 & $3/2^{+}$	&   								& -265.(4) 				& -0.163(40)[199]  \\
 38 & 3$^{+}$ 	& -985.9(4)[34] 	&   							& -0.126(3)[177] \\
    &  					&   								& -127.0(53) 			& -0.140(51)[174] \\
 39 & 3/2$^{+}$ & -862.5(9)[30] 	&   							& -0.037(8)[153]  \\
    &  					&   								&  	0 						& -0.082(15)[151]  \\
 40 & 4$^{-}$ 	&   								& 125.6(3)		   	& -0.066(16)[129]  \\
 41 & 3/2$^{+}$ &   								& 235.3(8)		   	&  0.036(17)[108] \\
 42 & 2$^{-}$ 	& -506.7(7)[17] 	&   							&  0.034(6)[89] \\
   	&   				&   								& 351.7(19)				&  0.026(23)[88]  \\
 43 & 3/2$^{+}$ &   								& 459.0(12)				&  0.049(19)[69] \\
 44 & 2$^{-}$ 	& -292.1(5)[10]  	&   							&  0.036(5)[51]  \\
  	&   				&   								& 564.3(14)				&  0.047(20)[50] \\
 45 & 3/2$^{+}$ &   								& 661.7(16)				&  0.072(21)[33]  \\
 46 & 2$^{-}$ 	& -91.6(5)[3]  	&   							& -0.002(4)[16] \\
  	&   				&   								& 762.8(15)				&  0.026(21)[16] \\
 47 & 1/2$^{+}$ &  0 								& 857.5(17)			  &  0 \\
 48 & 1$^{-}$ 	& 67.9(4)[3] 		&   							&  0.186(3)[16]  \\
 49 & 1/2$^{+}$ & 135.3(5)[6] 		&   							&  0.342(4)[32]  \\
 50 & 0$^{-}$ 	& 206.5(9)[9] 		&   							&  0.434(8)[47]  \\
 51 & 3/2$^{+}$ & 273.2(14)[11] 	&   							&  0.538(13)[61] \\
 \end{tabular}
\end{table}

\begin{figure}[htb]
\begin{center}
  \includegraphics[width=0.88\linewidth]{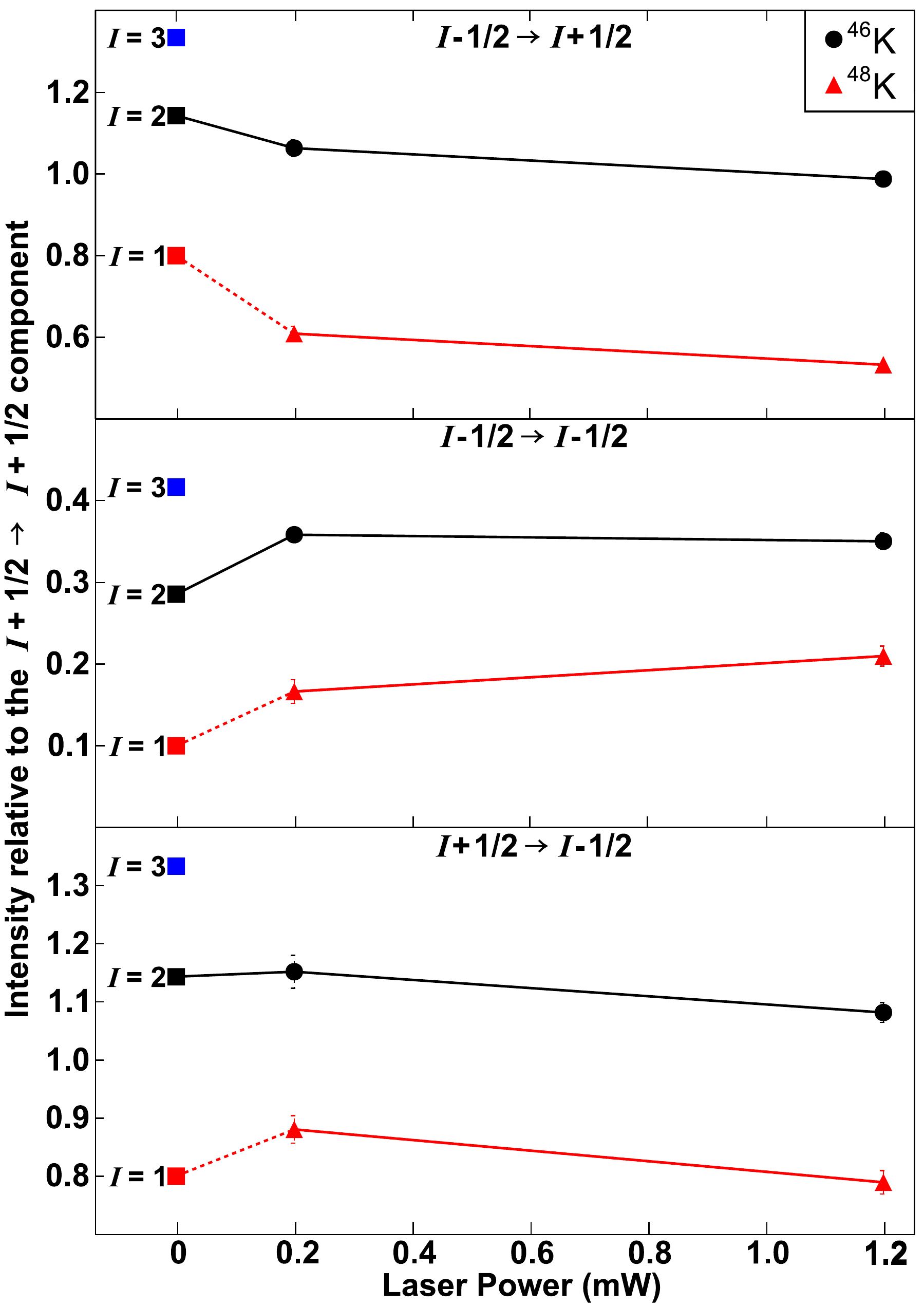}%
\end{center}
 \caption{(Color online) Intensities of the hfs components of $^{46}$K and $^{48}$K relative to the $I+1/2 \rightarrow I+1/2$ component. Data points at $0$\,mW laser power are expected intensities (Racah intensities) for different nuclear spin values. See text for details.\label{46K-48K}}
 \end{figure}

\begin{figure}[htb]
	\begin{center}
  \includegraphics[width=0.98\linewidth]{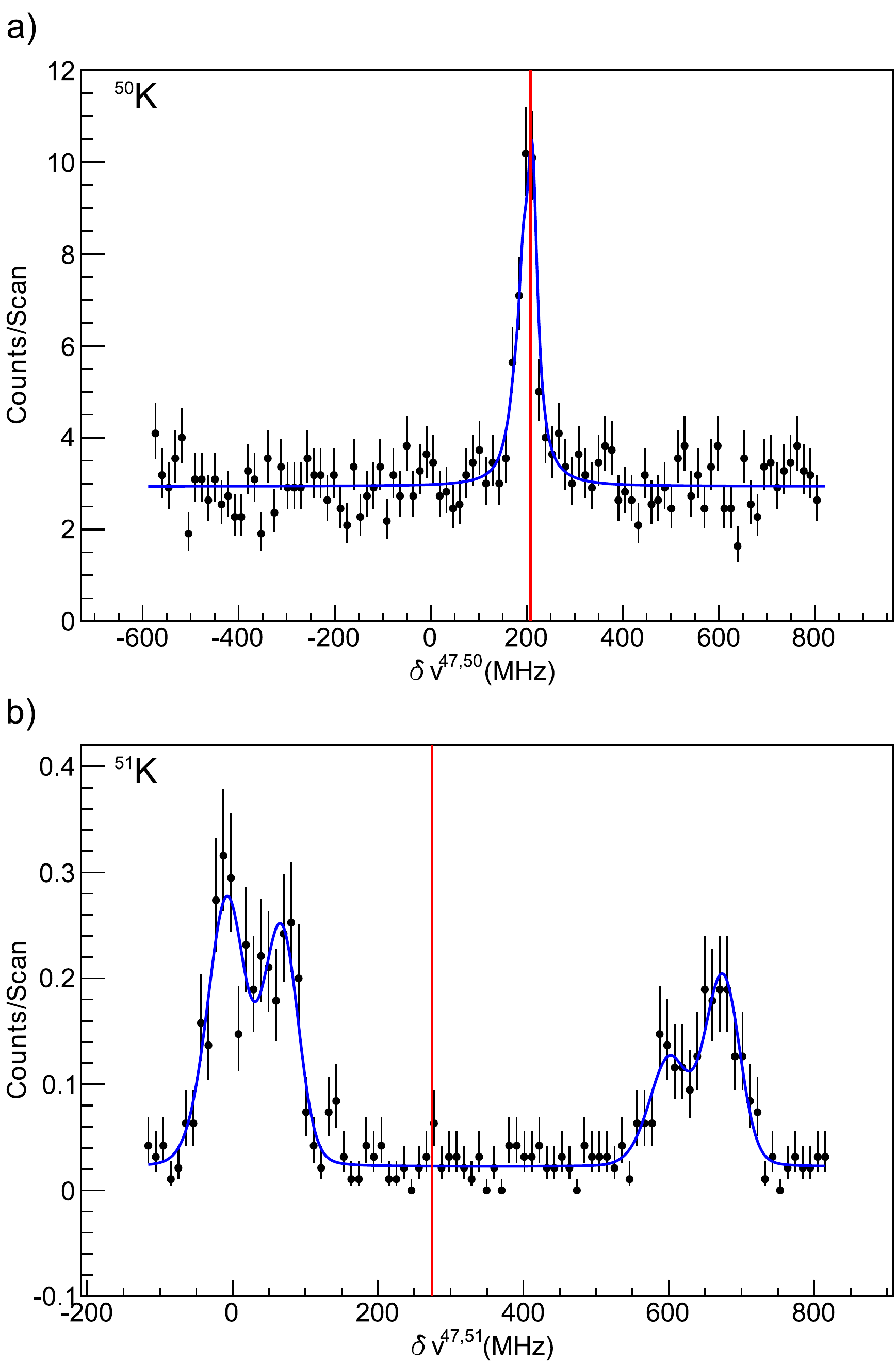}%
  \caption{(Color online) Optical spectra of $^{50}$K a) and $^{51}$K b) as in Fig.\ref{48K}\label{50K}}
	\end{center}
\end{figure}

For $^{48}$K our spin assignment is based on the different relative intensities of the hfs components for $^{46}$K ($I=2$) and $^{48}$K. In Fig.~\ref{46K-48K} we show the intensities for the individual hfs components of $^{46}$K and $^{48}$K relative to the $I+1/2 \rightarrow I+1/2$ (see Fig.~\ref{48K}) component as a function of laser power. Data points are taken at $0.2$\,mW and $1.2$\,mW, the relative intensities at $0$\,mW represent the theoretical Racah intensities \cite{Magnante1969}. The hfs of both isotopes have been recorded under identical experimental conditions and the only property that can cause a difference in the relative intensities is the nuclear spin. The relative intensities of $^{46}$K and $^{48}$K are significantly different and therefore the spins cannot be the same. From the measured intensity ratios and their extrapolation to zero laser intensity, compared to Racah values, it is clear that the $^{48}$K spin has to be $I=1$. This supports the $1^{-}$ assignment recently proposed by \citet{Krolas11} and excludes the $2^{-}$ adopted value in the 2006 nuclear data sheets \cite{Burrows06} . 

The hfs of $^{50}$K shows only one peak in the spectrum, see plot a) in Fig.~\ref{50K}. No fluorescence was observed in a broad scan of $1.4$\,GHz around the single peak. This is only possible for a ground-state spin of $I=0$, thus supporting the $0^{-}$ assignment by \citet{Baumann98} and excluding the $1^-$ configuration proposed by \citet{Crawford09}. 

For the determination of the ground-state spin of $^{51}$K the situation is more complex than in $^{48}$K. Spin $I=1/2$ can be excluded since four, and not three, peaks are visible in the recorded spectra. Figure \ref{50K} b) shows the hyperfine spectrum of $^{51}$K with a fit assuming a spin of $I=3/2$, which yields a good agreement with the data. 
However, $\chi^2$-analysis alone can not exclude the possibility of spin $5/2$. In our previous work \cite{Papuga12} spin $3/2$ was assigned mainly on the basis of the magnetic moment. To remove any ambiguity we consider here the relative line intensities, as in the case of $^{48}$K. Due to the exoticity of $^{51}$K, spectra were recorded only at the optimal laser power of $1.2$\,mW, hence a direct comparison with a particular isotope (e.g. $^{39}$K) for different laser powers is not possible. The solution is to compare the relative intensities of the hyperfine components of $^{51}$K with those of several other isotopes at $P=0$\,mW (Racah intensities) and $P=1.2$\,mW. 
\begin{figure*}[t]
\begin{center}
\includegraphics[width=\linewidth]{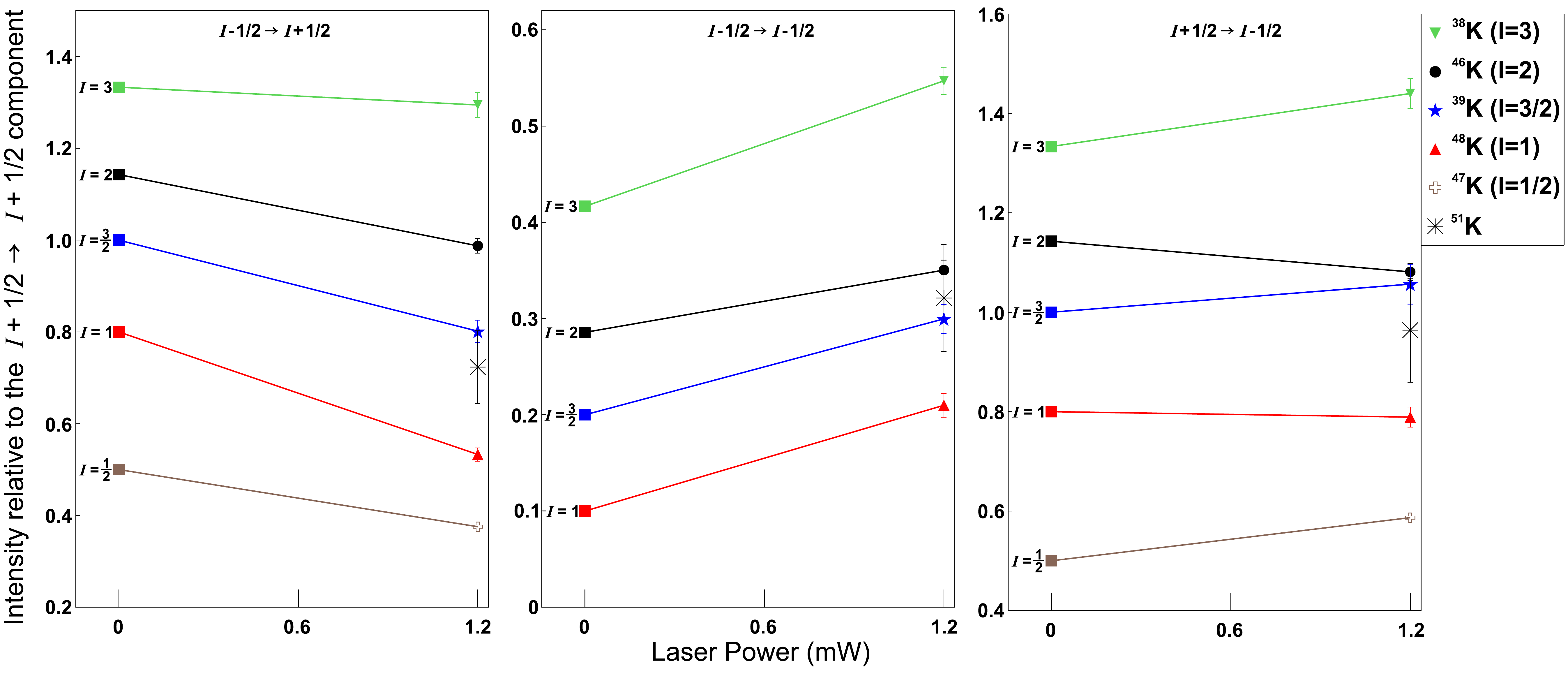}
\end{center}
 \caption{(Color online) Spin determination of $^{51}$K based on the intensities of the hyperfine components of $^{38,39,46,47,48,51}$K relative to the $I+1/2 \rightarrow I+1/2$ component. Theoretical intensity ratios for the corresponding nuclear spins are plotted at $0$\,mW, experimental ratios were determined at $1.2$\,mW laser power. For details see text.}
 \label{fig:51K-spin}
\end{figure*}

\begin{figure}[htb]
\begin{center}
\includegraphics[width=0.98\linewidth]{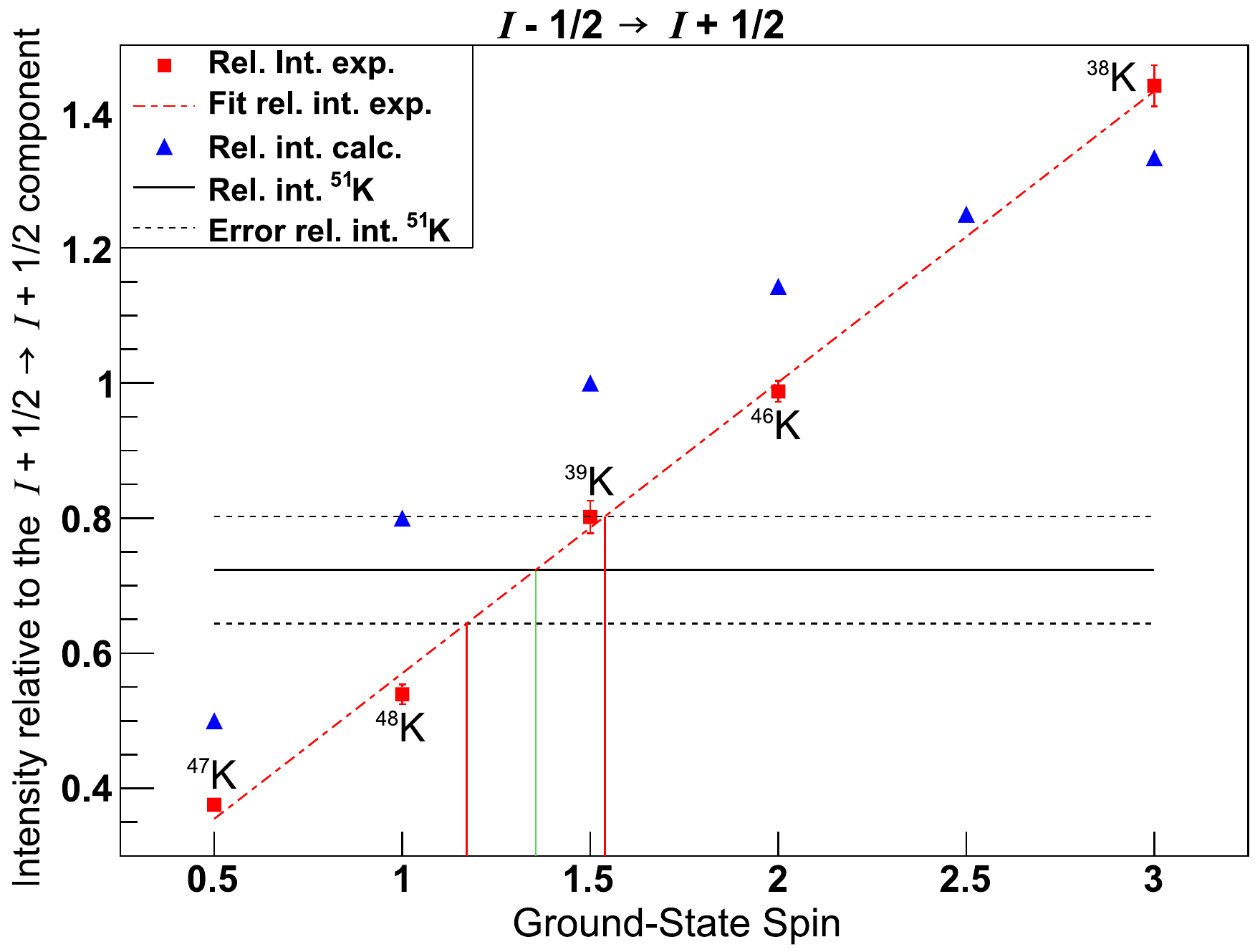}
\end{center}
 \caption{(Color online) Intensities of the hyperfine component $I-1/2 \rightarrow I+1/2$ of $^{38,39,46,47,48,51}$K relative to the  $I+1/2 \rightarrow I+1/2$ component as a function of ground-state spin. For details see text.}
 \label{fig:51K-spin-2nd}
\end{figure}

Figure \ref{fig:51K-spin} shows these intensities of $^{38,39,46,47,48,51}$K relative to the $I+1/2 \rightarrow I+1/2$ component as a function of laser power. Similar to Fig.~\ref{46K-48K}, the data points are connected to their respective value at $P=0$\,mW with a solid line. The intensities measured for $^{51}$K are denoted by asterisks. Within the error bars the relative intensities of $^{51}$K agree with those for a ground-state spin of $I=3/2$, corresponding to the spin of $^{39}$K.
To define a confidence level for the spin determination the observed relative intensities were plotted against the spin.
Figure \ref{fig:51K-spin-2nd} shows the intensities of the hyperfine component $I-1/2 \rightarrow I+1/2$ of $^{38,39,46,47,48,51}$K (left plot of Fig.~\ref{fig:51K-spin}) relative to the  $I+1/2 \rightarrow I+1/2$ component. These relative intensities of $^{38,39,46,47,48}$K were fitted with a linear function, represented by the dot-dashed line.
The relative intensity determined for $^{51}$K is given as a horizontal line shown with its error band (dashed lines). The intersection of the relative intensity of $^{51}$K with the fitted line projected on the spin axis defines $I=3/2$ as the ground-state spin. 
The spin $5/2$ (or higher) is more than $2 \sigma$ away, and therefore can be ruled out with a confidence of $95$\,$\%$. The plots for the two remaining intensity ratios from Fig.~\ref{fig:51K-spin} confirm this level of confidence. As a result, we determine the ground-state spin of $^{51}$K to be $I=3/2$, which supports the $3/2^+$ assignment made by \citet{Perrot06}.

With the measured spins for $^{48,50}$K, along with the magnetic moment of $^{48}$K, the discussion on the evolution of the $\pi d_{3/2}$ orbital when filling the $\nu p_{3/2}$ orbital can now be completed. The earlier reported spins and magnetic moments of $^{49}$K and $^{51}$K revealed a re-inversion of the $\pi s_{1/2}$ and $\pi d_{3/2}$ levels back to their normal ordering, as neutrons are added from $N=30$ to $N=32$ \cite{Papuga12}. A comparison of the measured observables to large-scale shell-model calculations has shown that like in $^{47}$K ($N=28$), the $\pi s_{1/2}^{-1}$ hole configuration is the dominant component in the $^{49}$K ground state wave function, although its magnetic moment suggests that the wave function has a significant admixture of a $\pi d_{3/2}^{-1}$ hole configuration. In $^{51}$K the wave function is dominated by the normal $\pi d_{3/2}^{-1}$ hole and from its magnetic moment a rather pure single particle wave function can be inferred. The wave functions of the odd-odd $^{48,50}$K isotopes have been investigated using the same effective shell-model calculations. A comparison to the experimental data along with a discussion on the occupation of the proton orbitals at $N=29$ and $N=31$ will be presented in a forthcoming paper (\citet{Papuga13}). 

The measured isotope shifts relative to $^{47}$K and literature values referenced to $^{39}$K \cite{Touchard1982,Behr1997} are given in Table \ref{Results}, where statistical errors are given in round brackets. The systematic errors, arising mainly from the uncertainty of Doppler shifts depending on the beam energy, are given in square brackets. 

The isotope shift between isotopes with atomic numbers $A, A'$ is related to the change in the nuclear mean square charge radii through:
\begin{equation}
\delta \nu ^{A,A'}=\nu^{A'}-\nu^{A}=K\frac{m_{A'}-m_{A}}{m_{A'}m_{A}}+F \delta \langle r^{2}\rangle^{A,A'}
\label{deltanu}
\end{equation}
with $\nu^{A}$ and $\nu^{A'}$	 representing the transition frequencies with respect to the fine structure levels, $K=K_{\text{NMS}}+K_{\text{SMS}}$ the sum of the normal and the specific mass shift, $m_{A}$ and $m_{A'}$ the atomic masses, $F$ the electronic factor, and $\delta \langle r^{2}\rangle^{A,A'}=\langle r^{2}\rangle^{A'}-\langle r^{2}\rangle^{A}$ the change in the nuclear mean square charge radius.
For the extraction of $\delta \langle r^{2}\rangle^{A,A'}$ from the measured isotope shifts, the specific mass shift ($K_{\text{SMS}}=-15.4(3.8)\,$GHz\,u) and the electronic factor ($F= -110(3)$\,MHz\,fm$^{-2}$) calculated in \cite{Martensson-Pendrill1990} were used. The normal mass shift ($K_{\text{NMS}} =\nu^{47}$m$_{e}$) was calculated to $K_{\text{NMS}} =213.55$\,GHz\,u using the D1-frequency of $^{39}$K measured in \cite{Falke2006}. The masses for $^{37}$K-$^{51}$K were taken from \cite{AME2012}. The calculated values of $\delta \langle r^{2}\rangle^{47,A}$ are shown in Table \ref{Results}. For the isotope shifts from \cite{Touchard1982, Behr1997} the same procedure was used to calculate the $\delta \langle r^{2}\rangle^{47,A}$ in order to obtain a consistent set of $\delta \langle r^{2}\rangle^{47,A}$ over the entire potassium chain.
The statistical error on $\delta \langle r^{2}\rangle^{47,A}$ (round brackets) originates from the statistical error on the isotope shift whilst the systematic error (square brackets) is dominated by the uncertainty on $K_{\text{SMS}}$ and is correlated for all isotopes.

\begin{figure}[htb]
\begin{center}
 \includegraphics[width=0.98\linewidth]{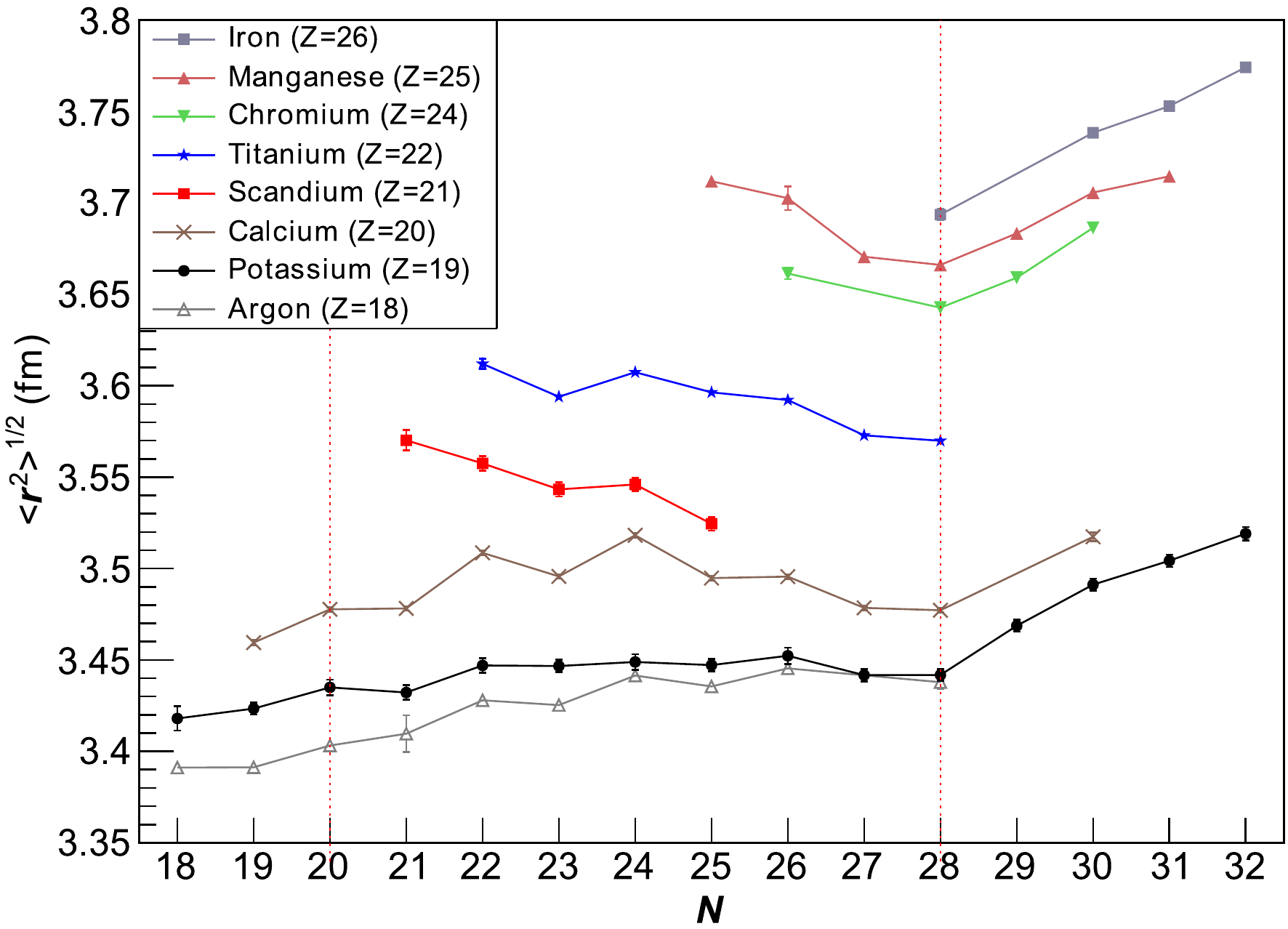}%
 \caption{(Color online) Root mean square nuclear charge radii versus neutron number for the isotopes of Ar, K, Ca, Sc, Ti, Cr, Mn and Fe. Data for K results from this work. \label{radii}}
\end{center} 
\end{figure}

In Fig.~\ref{radii} we compare the root mean square (rms) charge radii $\langle r^{2}\rangle^{1/2}$ of argon \cite{Blaum2008}, calcium \cite{Vermeeren92}, scandium \cite{Avgoulea11}, titanium \cite{Gangrsky04}, chromium \cite{Angeli04}, manganese \cite{Charlwood10}, iron \cite{Benton97}, and potassium from this work. The $\langle r^{2}\rangle^{1/2}$ values have been obtained by using the originally published changes in mean square charge radii and absolute reference radii from the compilation of Fricke and Heilig \cite{Fricke2004}.
The trend of the isotopic chains below $N=28$ has been discussed in \cite{Blaum2008, Avgoulea11}. In this region the behavior of the rms charge radii displays a surprisingly strong dependence on $Z$ (Fig.~\ref{radii}) with a general increase as a function of $N$ for Ar developing into a dominantly parabolic behavior for Ca and then progressing towards the anomalous downward sloping trends in Sc and Ti. Such dramatic variations of the radii are not observed in the regions around $N=50,82$ and $126$ \cite{Angeli09, Otten89}.

To date, no single theoretical model has fully described the strongly $Z$-dependent behavior of radii across $Z=20$ \cite{Avgoulea11}, although a variety of approaches have shown some success in describing the observed trends for specific isotopic chains. 

\begin{figure}[htb]
 \begin{center}
 \includegraphics[ width=0.98\linewidth]{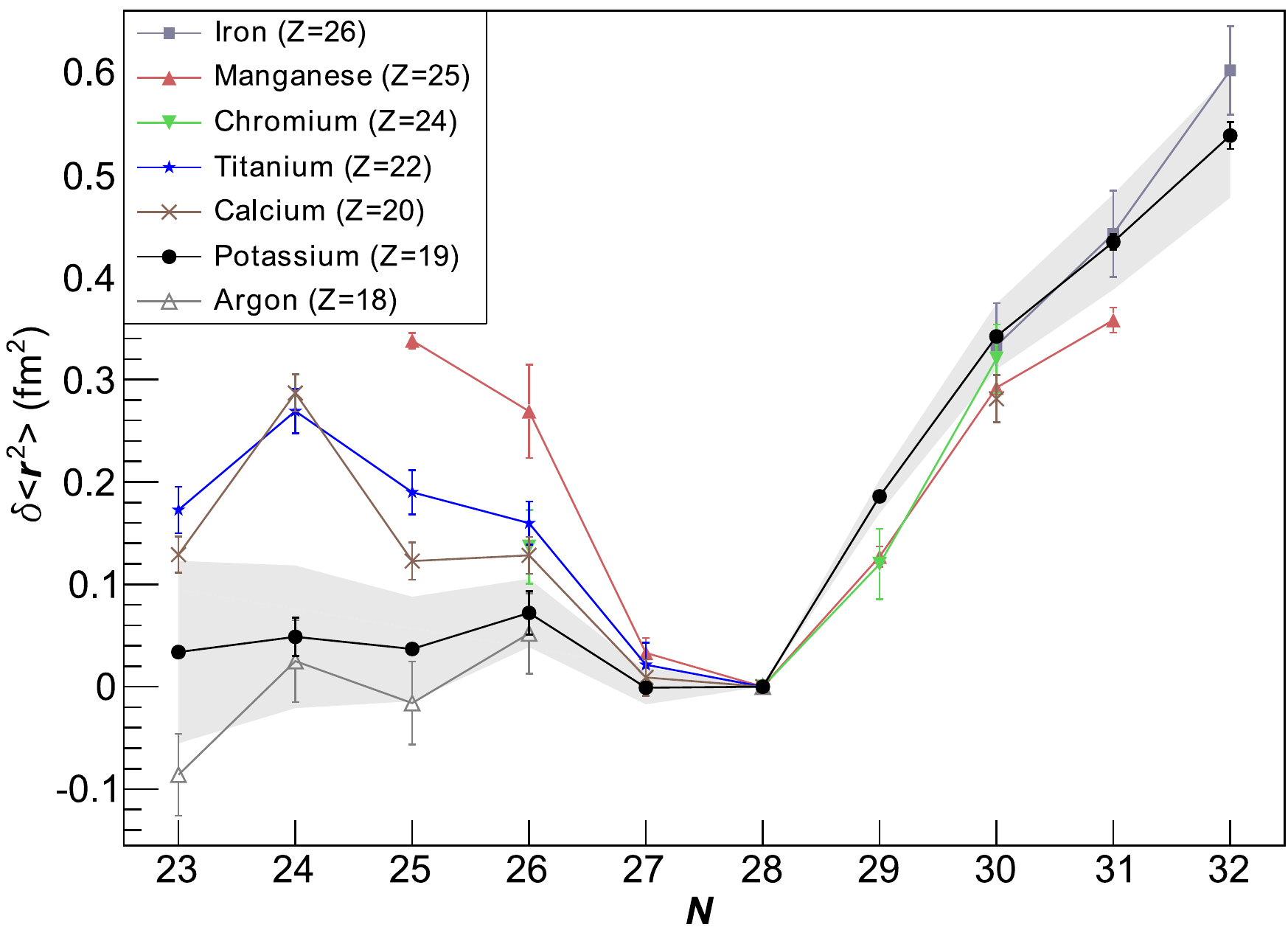}%
 \caption{(Color online) Difference in mean square charge radii relative to $N=28$ versus neutron number for the isotopes of Ar, K, Ca, Ti, Cr, Mn and Fe. \label{dradii}}
 \end{center}
 \end{figure}

To discuss the new results for $N>28$ we plot the $\delta \langle r^{2}\rangle$ of potassium given in Table \ref{Results} from $N=23$ to $N=32$ together with those of Ar, Mn, Ca, Ti, Cr and Fe referenced to the $N=28$ shell closure, see Fig.~\ref{dradii}. The correlated systematic error in the potassium data is represented by the gray shaded area surrounding the curve.
Looking at the $\delta \langle r^{2}\rangle$ curves one can see the above discussed $Z$ dependence of the radii in a broad structure below $N=28$. Common to all elements is the strong shell-closure effect at $N=28$. Above $N=28$ the $\delta \langle r^{2}\rangle$ values show a steep increasing slope, which is similar for all the elements and thus nearly independent of the number of protons. Below $N=28$ the $\delta \langle r^{2}\rangle$ values show large differences between the elements illustrating the atypical $Z$ dependence below $N=28$.

What changes at $N=28$? Up to $N=28$ the protons and neutrons in the calcium region ($18\leq Z\leq26$) occupy the same orbitals ($sd$ and $f_{7/2}$) resulting in a complex interplay of proton and neutron configurations. This fact is underlined by the shell-model calculations of \citet{Caurier01}, where the characteristic radii trend (see Fig.~2. in \cite{Caurier01}) in calcium between $N=20$ and $N=28$ is reproduced qualitatively by allowing multi-proton - multi-neutron excitations from the $sd$ to the $fp$ shell. Above $N=28$ the neutrons fill the $p_{3/2}$ orbital. Here the changes in mean square radii show little or no dependence on $Z$, indicating that charge radii are purely driven by a common and collective polarization of the proton distribution by the neutrons in the $p_{3/2}$ orbital with little or no dependence on the specific proton configuration at the Fermi surface. This absence of strong  proton configuration dependence is further emphasized by the consistency of the $^{47,49}$K radii, having an inverted ground-state proton configuration, with the radii of the remaining K isotopes as well as all other measured radii in the region. 
Furthermore the changes in the mean square radii of potassium show no indication of a shell closure at $N=32$, as discussed in \cite{Honma2005,Gade2008,Sorlin08}. Theoretical and experimental evidence of a shell effect at $N=32$ exists for Ca, Ti and Cr from the systematics of the first $2^+$ state energies $E_{x}(2^{+}_{1})$ or two-neutron separation energies $S_{2n}$ with the strongest effect in Ca \cite{Honma2005,Sorlin08,Wienholtz2013a, Steppenbeck13}.

Looking at the branch of $\delta \langle r^{2}\rangle$ of K above $N=28$ one can see that the point for $^{48}$K ($N=29$) deviates from the general odd-even behavior in the K chain as well as the great majority of all nuclei.  This general behavior of "normal" odd-even staggering is expressed by a smaller radius of the odd-$N$ compared to the neighboring even-$N$ isotopes.  Obviously the value for $^{48}$K is larger, corresponding to an "inverted" odd-even staggering. This feature might be related to the particular nature of the ground state wave function of $^{48}$K. Shell- model calculations \cite{Papuga13} show that the ground states of $^{47}$K and $^{49}$K are dominated by an $s_{1/2}$ configuration, while for $^{48}$K the dominant part comes from $d_{3/2}$. 

 \begin{figure}[htb]
 \begin{center}
 \includegraphics[ width=0.98\linewidth]{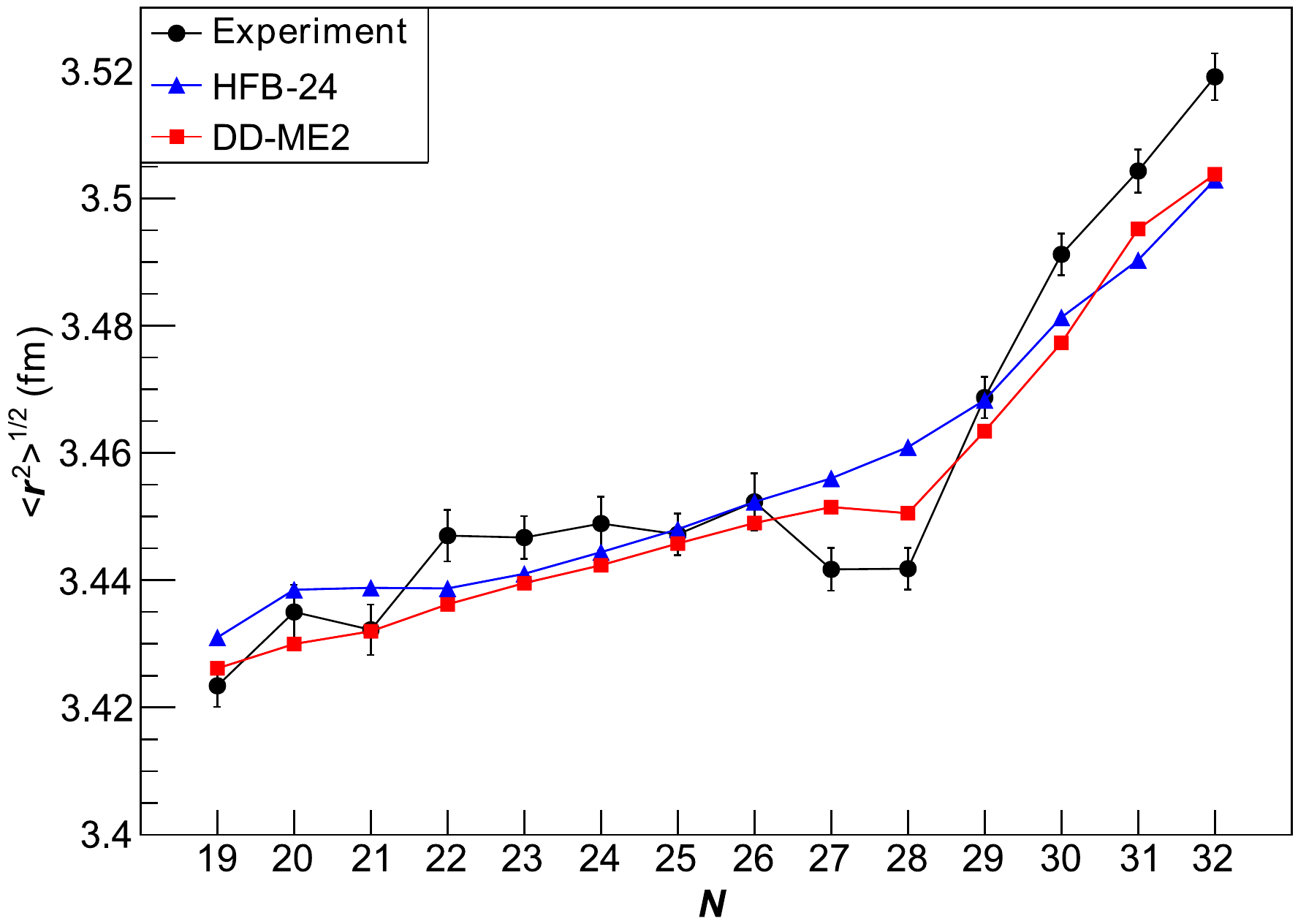}%
 \caption{(Color online) rms nuclear charge radii versus neutron number (black dots) compared to theoretical models HFB-24 (blue triangles) and DD-ME2 (red squares). \label{radii-theory}}
 \end{center}
 \end{figure}
The results of the rms charge radii are finally compared to theoretical calculations in the framework of the mean field (MF) model. Both the non-relativistic and relativistic MF approaches are considered because they are known to give rise to different descriptions of the spin-orbit field. The non-relativistic MF traditionally makes use of a phenomenological two-body spin-orbit term in the Skyrme force with a single parameter adjusted to experimental data, such as single-particle level splittings. The form of the spin-orbit term is, however, chosen purely by simplicity. In contrast, in the relativistic MF, the spin-orbit field emerges as a consequence of the underlying Lorentz invariance, and its form does not have to be postulated a priori. This approach, for the first time, led to a proper description of the generally observed charge-radii kinks at magic neutron numbers for the example of Pb \cite{sharma93}. As shown in Fig.~\ref{radii-theory}, such a kink is also predicted by the relativistic MF calculation, based on the DD-ME2 interaction \cite{lala05}, for the K isotopic chain, though the effect is less pronounced than found in our measurement. In contrast, none of the standard Skyrme forces with the traditional spin-orbit term gives rise to a proper description of the kinks. Such a behavior is illustrated in Fig.~\ref{radii-theory}, where the rms nuclear charge radii calculated with the  Skyrme-HFB  model \cite{Goriely2013}, HFB-24, are compared with experimental data. The parameters of the Skyrme interaction (BSk24) were in this case determined primarily by fitting measured nuclear masses and the properties of infinite nuclear matter.
It has been shown that a generalization of the Skyrme interaction \cite{rein95,sharma95} should also be able to map the relativistic spin-orbit field and improve the description of charge radii across the shell closures. 

It should be mentioned that neither the relativistic nor the non-relativistic MF models reproduce the parabolic shape between $N=20$ and $28$, nor the odd-even staggering. In particular, both calculations make use of the equal filling approximation and violate the time-reversibility in the treatment of odd nuclei, and consequently are not adequate for a proper description of the odd-even effect.

The hyperfine structure and isotope-shift measurements on $^{38,39,42,44,46\text{-}51}$K performed in the present work offer an insight into the development of the mean square charge radii in potassium beyond the $N=28$ shell closure. The now accessible range of K isotopes with neutron numbers $N=18$-$32$, including the previous data on $^{37\text{-}47}$K \cite{Touchard1982,Behr1997}, covers the full $f_{7/2}$ and $p_{3/2}$ orbitals.
Our measurement contributes substantially to a systematic investigation of mean square nuclear charge radii in the calcium region beyond the $N=28$ shell closure. The most striking effect is the difference in the behavior of the rms charge radii below and above $N=28$.
In addition the measured spins of $^{48,50}$K together with the spins reported by \citet{Papuga12} allow now assigning spins and parities of excited levels in these isotopes based on a firm ground and resolve the inconsistency between earlier reported data.

This work was supported by the Max-Planck-Society, BMBF (05 P12 RDCIC), the Belgian IAP-projects P6/23 and P7/10, the FWO-Vlaanderen, the NSF grant PHY-1068217 RC 100629, and the EU Seventh Framework through ENSAR (no. 262010).
We thank the ISOLDE technical group for their support and assistance.



%% References


\bibliographystyle{model1a-num-names}
\bibliography{K-Paper}


\end{document}